\documentclass[twocolumn,showpacs,preprintnumbers,amsmath,amssymb]{revtex4}
\usepackage{graphicx}
\usepackage{dcolumn}
\usepackage{bm}
\begin{document}
\title{Controlled dynamics of qubits in the presence of decoherence}
\author{D. D. Bhaktavatsala Rao}
\affiliation{Department of Physics, Indian Institute of Technology, Kanpur-208016,
INDIA}
\email{ddbrao@iitk.ac.in}
\date{\today}
\begin{abstract}
An exactly solvable model for the decoherence of one and two-qubit states interacting with a spin-bath, in the presence of a time-dependent magnetic field is studied. 
The magnetic field is static along $\hat{z}$ direction and oscillatory in the transverse plane.
The transition probability and Rabi oscillations between the spin-states of a single qubit is shown to depend on the size of bath, the distribution of qubit-bath couplings and the initial bath polarization. In contrast to the fast Gaussian decay for short times, the polarization of the qubit shows an oscillatory power-law decay for long times. The loss of entanglement for the maximally entangled two-qubit states, can be controlled by tuning the frequency of the rotating field. The decay rates of entanglement and purity  for all the Bell-states are same
 when the qubits are non-interacting, and different when they are interacting.

\end{abstract} 
\maketitle
\section{Introduction}
Solid-state spin systems have emerged to be promising candidates for quantum computation \cite{loss1,kane,petta1,awaschlom} and communication \cite{bose1,subbu}, as they offer the possibility for tunable device parameters and scalability.
However, their experimental realization requires control over the strong environmental decoherence. 
In most of the solid-state systems the interaction of qubits with the environmental spins is recognized to be a major source of decoherence \cite{veronica,stamp1}. For example in the case of quantum dots the hyperfine interaction of the electron-spin with the nuclear spins is the dominant mechanism of decoherence \cite{loss2,petta1}. In the case of SQUIDS the macroscopic quantum coherence is shown to be strongly suppressed because by a spin-bath consisting of nuclear paramagnetic spins and charge defects \cite{stamp}.

A well-studied model for decoherence is a central-spin system coupled to $N$ non-interacting spins\cite{breuer,bose,hamdouni,zurek,durga1,durga2,stamp1,dobro}. An exact evaluation of dynamics for the central-spins is obtained for special cases, where there is uniform coupling with the bath-spins \cite{breuer,hamdouni,durga1}, or a special choice of initial states\cite{loss3}, or the system-bath interaction is a simple interaction between the $z$ component of the spins\cite{zurek}.
Each such cases studied could explain the experimentally observed results in some solid-state devices \cite{loss2,durga1}.
In most of these studies the central-spin system either experiences a large static field or zero field. But, an important requirement for quantum computing is control, which can be either optical\cite{optcont,exp1} or through the time-dependent magnetic fields\cite{koppens}. In the case of single qubits, the control field is an oscillatory magnetic field, which can preform gate operations involving arbitrary spin rotations. On the other hand for two-qubits the exchange
interaction between the qubits is the control field, for the implementation of the controlled not gate. Recently, coherent spin rotations of a single electron
were demonstrated in a double quantum dot\cite{koppens}. 
Petta et.al., \cite{petta1} have implemented the CNOT gate or the $\sqrt{SWAP}$ gate, 
where the exchange interaction is controlled by gate voltages. 
In addition to the control of decoherence of the quantum state, it is equally important to control  the loss of entanglement in multi-qubit systems. For example a maximally entangled state shared between two remote sites looses its entanglement due to the local environments \cite{durga2}. In such cases, it is important to know whether a local control could slow-down the entanglement loss, thereby increasing the fidelity for teleportation.

 In this work we shall address the problem of decoherence and entanglement loss in the presence of a control field: a time-dependent magnetic field. 
We solve the dynamics exactly for the cases of a central-spin system consisting of one and two qubits. In the case of single-qubit the focus will be on the decay of Rabi-oscillations, whereas for the two-qubits it would be on the loss of entanglement. 
\section{Model}
We consider a central spin system consisting of one or more interacting qubits, coupled
to a bath (environment) of $N$ spin-$1/2$ particles. The Hamiltonian for this system is given
by
\begin{eqnarray}
\label{eqn1}
\hspace{-10mm}
\mathcal{H} &=& 
\sum_i \left[\hbar\omega_0 S^z_i + \hbar\omega_1 \left(S^{+}_i{\rm e}^{i\omega t}+S^{-}_i{\rm e}^{-i\omega t}\right)\right] \nonumber \\
&&+ \sum_{ij}J_{ij}\vec{S}_i\cdot\vec{S}_j
+ \sum_{ij}g_{ij}S^z_i I^z_j.
\end{eqnarray}
The first two terms represent the self Hamiltonian of the system spins $(\vec{S}_i)$, and
the last term describes the system-bath interaction, where the bath-spins are labeled by
$\vec{I}_i$. The qubits are coupled to a common external magnetic field ${\bf B}=[B_1\cos{\omega t}, B_1\sin{\omega t}, B_0]$, and $\omega_{0,1} = eB_{0,1}/m$, represents the strengths of the fields along the longitudinal and transverse directions respectively. The transverse field (rotating field) oscillates with a frequency $\omega$. The exchange interaction between the qubits is represented by $J_{ij}$ and the qubit-bath couplings by $g_{ij}$. We have neglected the
self Hamiltonian for the bath, as it is assumed to be weak in comparison to that of the other interactions in the model. Such an assumption is valid, for example, in the case of quantum dots, where the qubits are electron spins and the bath particles are spin-$1/2$ nuclei. Since the magnetic moment of electron, is much large $\large{O}(10^3)$ in comparison to that of the nucleon, the interaction among the bath-spins and their coupling to external fields have a comparatively small effect on the dynamics. It was argued earlier by Koppens $et.al.,$ \cite{koppens1}, that for sufficiently large external magnetic fields i.e., 
$\omega_0 \gg \sum_{j}g_{ij}, \omega_1$, the system-bath interaction considered in the above Hamiltonian can replace the usual hyperfine interaction between the electron and nuclear spins.
Hence the above Hamiltonian can describe the dynamics of the electronic spin states in quantum dots, in the regime of strong magnetic fields. 

The purpose of the work is two fold, to study the effect of a time-dependent field on the decoherence and the entanglement loss of the qubits. First in Sec-II, we consider the case of a single central spin, and study the effect of bath-interaction on the Rabi oscillations between the
states $|\uparrow\rangle$ and $|\downarrow\rangle$. In detail we shall consider two-different distributions for the coupling strengths, $g_k$, viz., (i) $g_k = g$, $\forall$ $k$ and (ii) $g_k = g\exp(-\alpha k^2)$. The Gaussian distribution for coupling strengths is of relevance in the context of a quantum dot \cite{coish}. We also consider the effect of bath-polarizations on the qubit dynamics.
In Sec-III we study the cases of interacting and non-interacting spins coupled to the bath. An additional feature that the two-spin states have is the entanglement. We shall address the question of whether the loss of entanglement between the system spins can be controlled the oscillating frequency. For both the cases, results are obtained analytically.

\section{Single spin dynamics and Rabi oscillations}
In the case of a single central spin the Hamiltonian given in Eq.\ref{eqn1} gets simplified to
\begin{eqnarray}
\label{eqn2}
\mathcal{H} = 
\hbar\omega_0 S^z + \hbar\omega_1 \left(S^{+}{\rm e}^{-i\omega t}+S^{-}{\rm e}^{i\omega t}\right) + S^z\sum_{k}g_{k}I^z_k. \nonumber \\
\end{eqnarray}
The initial state of the qubit-bath system is uncorrelated and given by $
\rho(0) = \rho_S(0)\otimes\rho_B(0)$, where $\rho_S, \rho_B$ represent the density matrices of the qubit and the bath respectively. Throughout this paper we shall consider the initial state of the qubit to be $\rho_S(0)=|\uparrow\rangle\langle\uparrow|$ and the bath state
to be 
\begin{eqnarray}
\label{eqnb}
\rho_B(0)&=\rho_{B_1} \otimes \rho_{B_2} \cdots \otimes \rho_{B_N}, &~~ \rho_{B_i} = \frac{1}{2}\mathcal{I} + P_BI^z_{i}.
\end{eqnarray}
The bath spins are polarized along the $\hat{z}$ direction, and the value of their common polarization is denoted by $P_B$. For $P_B=\pm 1$, $\rho_B$ is pure, and completely unpolarized ($\rho_B = \frac{1}{2^N}\hat{\mathcal{I}}$), for $P_B=0$.

The evolution of the total-system is governed by the dynamical equation
\begin{eqnarray}
\label{eqnr}
\rho(t) = U(t)\rho_S(0)\otimes \rho_B(0) U^{\dagger}(t),
\end{eqnarray}
where $U(t) = \exp{(it\mathcal{H}/\hbar)}$ is the time-evolution operator.
The reduced matrix of the qubit spin is obtained by tracing over the bath degrees of freedom,
$\rho_S(t) = {\rm Tr}_B \rho(t)$.
Before a general expression of $U(t)$ for the full Hamiltonian is given, we shall consider few simple cases.
\subsection{Zero coupling, $g_k$ = 0}
This is the simplest and well known case studied in the context of Rabi oscillations.
Setting the system-bath interaction to zero, we are left with only the first term in the Hamiltonian (see Eq.\ref{eqn2}). The unitary operator is straightforward to calculate and is given by
\begin{eqnarray}
\label{eqn3}
{U}_q(t) =\hat{V}\left[ \cos\frac{t}{2} \Omega ~\hat{\mathcal{I}} + 2i\frac{\sin\frac{t}{2}\Omega}{\Omega} \left\lbrace {\Delta_0}S^z + \omega_1 S^x\right\rbrace \right]
\end{eqnarray}
In the above $\hat{V}$ is the transformation matrix connecting the lab frame and the rotating frame, given by
\begin{eqnarray}
V = \left[\begin{array}{cc}{\rm e}^{i\omega t/2} & 0 \\
                              0 & {\rm e}^{-i\omega t/2}\end{array}\right ].
\end{eqnarray}
The Rabi frequency $\Omega = \sqrt{\omega^2_1 + \Delta^2_0}$, and $\Delta_0 = \omega-\omega_0$.
The Rabi oscillations occur when $\Delta_0 = 0$, and the transition probability for the qubit state
to be $|\downarrow\rangle \langle\downarrow|$ at any later time is given by
\begin{eqnarray}
\label{eqn4}
P_{\downarrow}(t,\omega) \equiv \langle \downarrow | \rho_S(t)| \downarrow\rangle =  \frac{\omega^2_1}{\Omega^2}\sin^2\frac{t}{2} \Omega.
\end{eqnarray}
The frequency of Rabi oscillations depend on the strength of the transverse field ($\omega_1$).

\subsection{One spin bath}
This is the minimum dimension of the bath, and the simplest case of qubit-bath interaction.
In the Hamiltonian (see Eq.\ref{eqn2}) $g_k = 0, \forall~k\ne 1$.
The unitary operator for this case is also straightforward to calculate and is given by
\begin{eqnarray}
{U}(t) = {U}^{+}_q(t)|\uparrow\rangle_b\langle\uparrow| + {U}^{-}_q(t)|\downarrow\rangle_b\langle\downarrow|,
\end{eqnarray}
where
\begin{eqnarray}
\label{eqn5}
{U}^{\pm}_q(t) = \hat{V}\left[ \cos\frac{t}{2} \Omega_{\pm}~\hat{\mathcal{I}} + 2i\frac{\sin\frac{t}{2} \Omega_{\pm}}{\Omega_\pm} \left\lbrace {\Delta_\pm}S^z + \omega_1 S^x\right\rbrace \right].
\end{eqnarray}
The coefficients $\Delta_{\pm} = \omega-(\omega_0 \pm g_1/2)$, and $\Omega_{\pm} = \sqrt{\omega^2_1 + \Delta^2_{\pm}}$.

\begin{figure}[htb]
   \includegraphics[width=8.0cm]{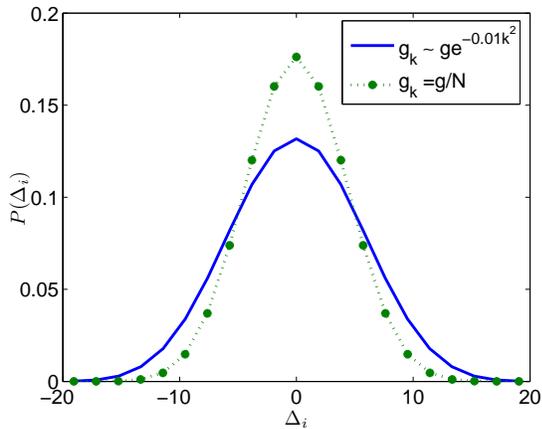}
 \caption{(Color online) Probability distribution of the resonant frequencies given in Eq.\ref{resf}. 
The bath consists of $N = 20$ spins. The distributions for the qubit-bath couplings are normalized such that $\sum_kg_k = g$, where $g/\hbar = 20$MHz. In plotting the above we have set $\omega = \omega_0$.
}
 \end{figure}
Depending on which state the bath spin is, the resonance can occur either at $\omega = \omega_0 + g_1/2$
or $\omega = \omega_0 - g_1/2$. For example when the initial state of the bath spin is $|\psi\rangle_B = \frac{1}{\sqrt{2}}[|\uparrow\rangle + |\downarrow\rangle] $,
the total state of the qubit-bath system can be written as
\begin{eqnarray}
|\psi\rangle = a_+(t)|\uparrow \uparrow\rangle + b_+(t)|\downarrow \uparrow\rangle +a_-(t)|\uparrow \downarrow\rangle + b_-(t)|\downarrow \downarrow\rangle. \nonumber \\
\end{eqnarray}
The cofficients 
\begin{eqnarray}
a_{\pm} &=& {\rm e}^{i\omega t/2}\left[\cos\frac{t}{2} \Omega_{\pm} + i\frac{\Delta_{\pm}}{\Omega}\sin\frac{t}{2} \Omega_{\pm}\right], \nonumber \\
b_{\pm} &=& i{\rm e}^{-i\omega t/2}\frac{\omega_1}{\Omega}\sin\frac{t}{2} \Omega_{\pm}.
\end{eqnarray}
The above state can be entangled if $|a_+b_- - a_-b_+|\ne0$.
Since the time-dependent coefficients depend on $\omega$, 
their entanglement can be controlled by tuning the oscillating frequency $\omega$ appropriately.  For $\omega = \omega_0-g_1/2$ and $g_1 \gg \omega_1$, then the above state at $t = \pi/\omega_1$, is close to the maximally entangled state $|\psi\rangle = \frac{1}{\sqrt{2}}[|\downarrow \uparrow\rangle +|\uparrow \downarrow\rangle]$ as $a_+ =0$, and $b_- \sim 0$ ($\omega_1 \ll\Omega_-$). On the other hand
when $\omega = \omega_0 + g_1/2$,
the qubit-bath state is close to the maximally entangled state
$|\psi\rangle = \frac{1}{\sqrt{2}}[|\uparrow \uparrow\rangle +|\downarrow \downarrow\rangle]$.
Note that the two maximally entangled states can be differentiated, by the knowing the value of $\omega$. Even though this could be used as a controlled entangling operator, or for the Bell-state discrimination, the resonance conditions are not good from the view point of the qubit, as it looses all its polarization. 

\begin{figure}[htb]
   \includegraphics[width=8.0cm]{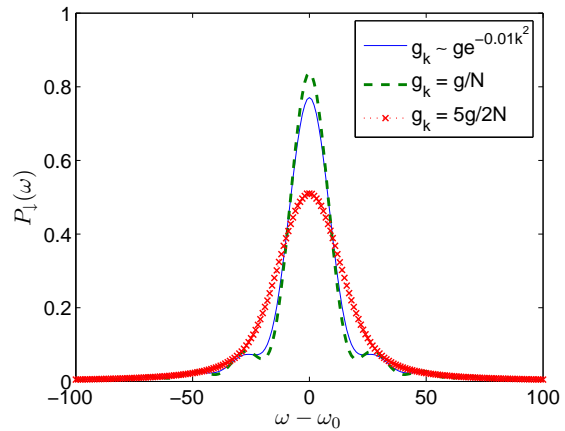}
 \caption{(Color online)Transition probability as a function of time. The initial and final states
for the transition are $|\uparrow\rangle$ and $|\downarrow\rangle$. The bath is unpolarized ($P_B=0$) consisting of $N=20$ spins. Two different distributions for the qubit-bath couplings are considered. The Gaussian distribution for $g_k$ is normalized such that $\sum_kg_k=g$. In plotting the above we have taken $\omega_0 = 10^3$, $\omega_1 = 10$ and $g/\hbar = 40$, which are in the units of MHz. }
 \end{figure}

\subsection{N spin bath}
We now go to the more general case of the bath. Since the Hamiltonian commutes with the total $I^z$ value of the bath, its is block diagonal in the $I^z$ basis. Hence, all it requires is to diagonalize a $2\times2$ matrix of the qubit Hamiltonian corresponding to each $I^z$ state.
The unitary operator can then be simply written as
\begin{eqnarray}
\label{eqn6}
{U}(t) = \sum_{i=1}^{2^N}{U}^i_q|i\rangle\langle i|,
\end{eqnarray}
where $|i\rangle$ are the basis states of the bath and the expression for ${U}^i_q$ is same as that given in Eq. \ref{eqn5}, except that the $\pm$ sign is replaced by the index $i$.
The Rabi frequency $\Omega_i = \sqrt{\omega^2_1 + \Delta^2_i}$, depends explicitly on the bath state where
\begin{figure}[htb]
   \includegraphics[width=8.0cm]{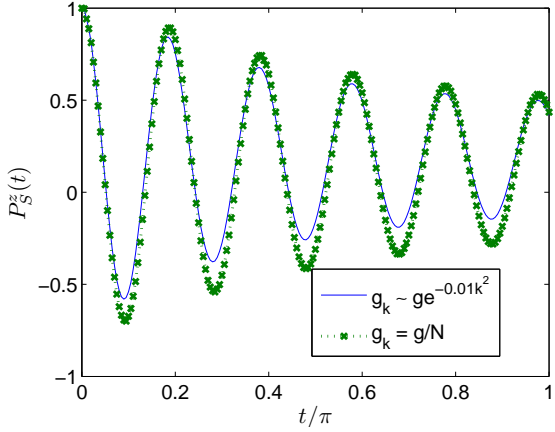}
 \caption{(Color online) The $z$-component of qubit polarization with time. The plot shows the decay of Rabi oscillations with time. The bath is unpolarized consisting of $N=20$ spins. Two different distributions for the qubit-bath couplings are considered, where $\sum_kg_k=g$ for both the distributions. In plotting the above we have taken $\omega_0 = 10^3$, $\omega_1 = 10$ and $g/\hbar = 40$, which are in the units of MHz}
 \end{figure}
\begin{eqnarray}
\label{resf}
\Delta_i = \omega-\left[\omega_0 + \langle i|\sum_k g_k I^z_k|i\rangle\right].
\end{eqnarray}
The frequency at which the resonance occurs is no more a single value but rather a distribution
determined by the coupling strengths $g_k$. 
For example when all $g_k=g$, the frequencies
$\Delta_i$ are distributed between $-gN/2 \le \Delta_i \le gN/2$. In the limit of large $N$,
this is a Gaussian distribution centered around $\Delta_i = \omega-\omega_0$. Hence, the dominant frequency at which the resonance occurs is still located at $\omega=\omega_0$. 
In Fig.1 we have plotted the distribution of frequencies $\mathcal{P}(\Delta_i)$ for two different distributions of the coupling strengths, (i)$g_k = g, \forall$ $k$, and
(ii)$g_k \sim g\exp(-0.01k^2)$. In both the cases, the distribution is centered around
$\omega_0$, implying that the resonance condition for Rabi oscillations is still $\omega = \omega_0$.

The reduced density matrix of the qubit can be evaluated from Eq.\ref{eqnr} by tracing over the bath degrees of freedom. Using the above obtained unitary operator, it can be written as
\begin{eqnarray}
\rho_S(t) = \sum_{i=1}^{2^N} {U}_i \rho_S(0) {U}^{\dagger}_i \langle i| \rho_B(0)|i\rangle.
\end{eqnarray}
For the initial bath state given in Eq.\ref{eqnb}, the above form gets simplified
to
\begin{eqnarray}
\label{eqnred}
\rho_S(t) =  \frac{1}{2^N}\sum_{i=1}^{2^N} \lambda_i {U}_i \rho_S(0) {U}^{\dagger}_i,
\end{eqnarray}
where $\lambda_i = (1+P)^{N/2+m_i}(1-P)^{N/2-m_i}$, and $m_i = 2 \langle i |\sum_k I^z_k |i\rangle$. For the initial state of the qubit $\rho_S(0)=|\uparrow\rangle\langle\uparrow|$, the transition probability to the spin down state is given by
\begin{eqnarray}
P_{\downarrow}(t,\omega) \equiv \langle\downarrow|\rho_S(t)|\uparrow\rangle = \frac{1}{2^N}\sum_i \lambda_i\frac{\omega^2_1}{\Omega^2_i}\sin^2\frac{t}{2}\Omega_i
\end{eqnarray}
\begin{figure}[htb]
   \includegraphics[width=8.0cm]{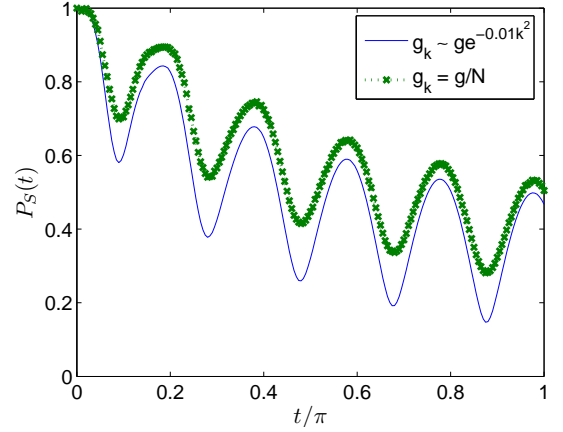}
 \caption{(Color online) Qubit polarization with time. The initial bath state is unpolarized consisting of $N=20$ spins. Two different distributions for the qubit-bath couplings are considered, where $\sum_kg_k=g$ for both the distributions. In plotting the above we have taken $\omega_0 = 10^3$, $\omega_1 = 10$ and $g/\hbar = 40$, which are in the units of mHz. The decoherence of one-qubit states ($1-{P}_S(t)^2$), can be inferred directly from the above figure.}
 \end{figure}

In Fig.2 we have plotted the transition probability $P_{\downarrow}(\omega)$, fixing the time at which the first maximum occurs i.e., $t=\pi/\omega_1$. The bath is completely unpolarized consisting of $N=20$ spins. Two different distributions for the coupling strengths $g_k$ are considered. Eventhough the peak of $P_{\downarrow}(\omega)$ occurs at $\omega = \omega_0$, 
the width of the distribution $P_{\downarrow}(\omega)$ depends upon the number of bath spins and the distribution of $g_k's$. Either with the increase of the bath-size increases or the qubit-bath coupling strength, the width of $P_{\downarrow}$ increases, and the probability for the transition to the down state decreases. Similarly if the distribution $g_k$ becomes narrower, the transition probability decreases, and reaches a minimum of $1/2$, when only one value of $g_k$ is non-zero. To see the decay of Rabi-oscillations we evaluate the qubit-polarizations as a function of time fixing the r.f frequency at $\omega = \omega_0$.

Using the expression for the reduced density matrix of the qubit (see Eq.\ref{eqnred}), it is straight forward to evaluate the time-dependent polarizations $P^i_S(t)\equiv 2{\rm Tr}(\rho_S(t)S^i)$, given by
\begin{eqnarray}
\label{poleqn}
P^z_S(t) &=& \frac{1}{2^N}\sum_{i=1}^{2^N} \lambda_i \left[1-2f^2_i(t) \right]\nonumber \\
P^x_S(t)&=&\frac{1}{2^{N-1}} \sum_{i=1}^{2^N} \lambda_i f_i(t)\left[\frac{\Delta_i}{\Omega_i}\cos\omega t \sin\frac{t}{2}\Omega_i+
\sin\omega t \cos\frac{t}{2}\Omega_i \right] \nonumber \\
P^y_S(t)&=&\frac{1}{2^{N-1}}\sum_{i=1}^{2^N} \lambda_i f_i(t)\left[\cos\omega t \cos\frac{t}{2}\Omega_i-\frac{\Delta_i}{\Omega_i}\sin\omega t \sin\frac{t}{2}\Omega_i\right] \nonumber \\
\end{eqnarray}
where $f_i(t) = \frac{\omega_1}{\Omega_i}\sin\frac{t}{2}\Omega_i$.
We now consider the cases of  polarized and unpolarized baths separately.

\subsubsection{Unpolarized bath}
If we set $P_B=0$ in Eq.\ref{eqnb}, then the resultant bath state is completely unpolarized, i.e.,
$\rho_B = \frac{1}{2^N}\hat{\mathcal{I}}$. Since the state has no polarizations $\lambda_i = 1$.
In Fig.3 we have plotted the variation of $z$ component of the polarization with time, for the r.f frequency $\omega$, set equal to $\omega_0$. The polarization has a damping behavior with time. 
For the case of uniform coupling strengths $g_k=g$, the above expression for the $\hat{z}$ component of the qubit-polarization gets simplified to
\begin{eqnarray}
P^z_S(t) = \frac{1}{2^N}\sum_{m=-N/2}^{N/2}C^N_{N/2-m}[1-2f^2_m(t)]
\end{eqnarray}
For large $N$, $C^N_{N/2-m}/2^N \approx \sqrt{\frac{2}{\pi N}}\exp(-2m^2/N)$ (central limit theorem). Replacing the summation by integration over the $m$ values the expression for $P^z_S(t)$ given above reduces to
\begin{eqnarray}
P^z_S(t) &=& \frac{\cos\left(\omega_1 t + \frac{1}{2}\tan^{-1}\gamma \omega t\right)}{(1+\gamma^2t^2)^{1/4}} \nonumber \\
&&+ \gamma\left(1 -\frac{\cos\left(\omega_1 t + \frac{3}{2}\tan^{-1}\gamma \omega t\right)}{(1+\gamma^2t^2)^{3/4}}\right),
\end{eqnarray}
where $\gamma = Ng^2/4\omega^2_1$.  One can see that there is an oscillatory power-law decay for long times. The behavior for $P^z(t)$ would be similar even when the distribution of $g_k$'s are from a Gaussian distribution. The polarization ${P}_S(t)$ plotted in Fig.4, gives the direct measure of decoherence $1-{P}_S(t)^2$ for the qubit.

\subsubsection{Polarized bath}
We now switch on the bath polarization. For $P_B=\pm 1$, the bath state is pure and the frequency at which the resonance occurs get shifted by $\pm gN/2$. For other values of polarization ($|P_B| < 1$)
the bath state is mixed and it is difficult to estimate the exact value of resonant frequency for a arbitrary distribution of $g_k's$. In Fig.5 we have plotted the transition probability as a function of $\omega$, when the coupling between the qubit and bath-spins are same i.e., $g_k = g$. The shift in the resonant frequency goes as $gPN/2$. 
This shift can be understood from the distribution of $m_i$ values. For non-zero $P_B$, and in the large $N$ limit, the eigenvalues of the bath operator $I^z$ are distributed as
\begin{eqnarray}
P(m) = \sqrt{\frac{2}{\pi N(1-P_B^2)}}\exp\left[-\frac{2(m-NP_B/2)^2}{N(1-P_B^2)}\right].
\end{eqnarray}
As the bath polarization increases, the oscillations of $P^z_S(t)$ becomes more stabilized between $\pm 1$, indicating perfect Rabi oscillations. But the resonant frequency is shifted away from $\omega_0$ by $gNP_B/2$.

\begin{figure}[htb]
   \includegraphics[width=8.0cm]{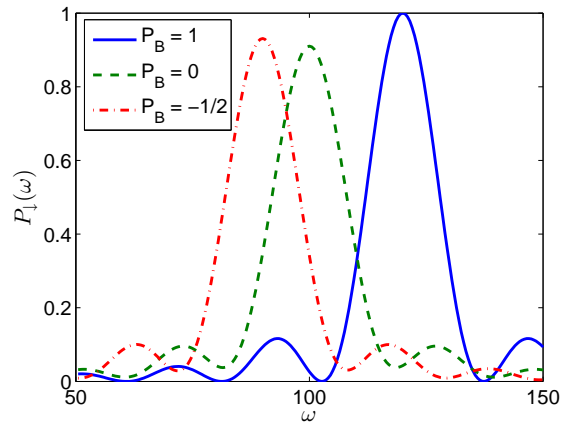}
 \caption{(Color online) The transition probability as a function of r.f frequency $\omega$.
Initial bath states with different polarizations are considered. The resonant frequency varies with the bath polarization as $\omega = \omega_0 + gNP_B/2$. In plotting the above we have taken
$N=20$, $\omega_0=100$, $\omega_1 = 10$ and $g/\hbar=1$. The frequencies are in the units of MHz.}
 \end{figure}

\section{Two-spin dynamics and controlled entanglement}
In the section we shall consider the case of a central spin system consisting of two spin-$1/2$ particles which can be either interacting or non-interacting. We consider both the cases. The Hamiltonian describing the two-qubit dynamics is given by (see Eq.\ref{eqn1}),
\begin{eqnarray}
\label{tqh}
\mathcal{H} &=& 
\left[\hbar\omega_0 S^z_{12}+ \hbar\omega_1 \left(S^{+}_{12}{\rm e}^{i\omega t}+S^{-}_{12}{\rm e}^{-i\omega t}\right)\right] \nonumber \\
&&+J\vec{S}_1\cdot\vec{S}_2+ \sum_{k}(g^{(1)}_{k}S^z_1+g^{(2)}_{k}S^z_2)I^z_k.
\end{eqnarray}
where $\vec{S}_{12} = \vec{S}_1 + \vec{S}_2$ represents the total spin of qubits.
Depending upon the common set of bath spins with which the qubits are interacting, we can either have a common or separate baths for the qubits. The total
spin of the two-qubit system $\vec{S}_{12} = \vec{S}_1+\vec{S}_2$ is invariant under time-evolution, only in the case of common spin environment. Hence for separate baths, we have transitions between the singlet and triplet sectors, which makes the dynamics more interesting. 
In case of quantum dots the overlap between the spatial wave functions of the
electrons, determines the common and separate nuclear environments for the electrons.
For strong overlap indicating a large value of $J$, both the electrons interact with the same set of neighboring nuclear spins, and on the other hand, when the two electrons are physically
apart, the overlap is quite small ($J$ small) indicating separate nuclear environments. 
Since $J$ can be controlled by gate voltages one can tune from one regime to the other.
In the case of two-qubits the main interest is to study the,
the swap operation between qubit states ($|\uparrow\downarrow\rangle \leftrightarrow |\downarrow\uparrow\rangle$), and the loss of entanglement for maximally entangled states. Because of the presence of a time-dependent transverse field the swap operation can controlled both by tuning the rotating frequency $\omega$ and the exchange interaction $J$. We use the concurrence measure $\mathcal{C}(t)$ \cite{woott}, to evaluate the entanglement of the two-qubit state.
Similar to the considerations made in Sec-I, we take the initial qubit-bath state to a
direct product $\rho(0) = \rho_{12}(0) \otimes \rho_B(0)$, where
$\rho_{12}$ can in general, be represented as
\begin{eqnarray}
\label{tqs}
\rho_{12}={\hat{\mathcal{I}}\over 4} + {1\over 2}\vec{P}_1\cdot\vec{S}_1 + {1\over 2}\vec{P}_2\cdot\vec{S}_2 + \sum_{m,n=1}^{3}\Pi^{mn}S^m_{1}S^n_{2}, \nonumber \\
\end{eqnarray}
where the vector polarizations are given by $\vec{P}_{1,2} \equiv 2Tr[\rho_{12}\vec{S}_{1,2}]$, and the
cartesian components of the tensor polarization are $\Pi^{mn}\equiv 4Tr[\rho_{12}S^m_1S^n_2]$.
We note that for a pure state ($\rho^2_{12}=\rho_{12}$), we have $P_1=P_2 \le 1$, and 
$P^2_1 + P^2_1 + \sum_{mn}(\Pi^{mn})^2 = 3$. The state of the bath is same as given in Eq.\ref{eqnb}.

We now evaluate the reduced density matrix of the two-qubit state, for the cases of common and separate nuclear baths.

\subsection{Common spin-bath for the qubits}
In this section we shall consider the case of a common spin environment for the qubits i.e., $g^{(1)}_k = g^{(2)}_k$ in the Eq.\ref{tqh}. Since, the total spin of the two-qubit system is conserved, the triplet and singlet sectors evolve separately. Similar to the analysis for single-qubit we start with no bath coupling and then switch on the qubit-bath interaction to study the reduced dynamics of the qubits. This would help to clearly identify the additional features that the bath-interaction can bring to the time-evolution of two qubit states.
\subsubsection{Zero bath coupling}
When there is no coupling with the bath, the time-evolution operator  can be found by diagonalizing the Hamiltonian in the four dimensional space, by making a transformation to the rotating frame.
The time-evolution operator can then be constructed straight forwardly giving
\begin{eqnarray}
\label{tqutop}
\hspace{-10mm}
U_{12}(t) &=& \hat{V}_{12}{\rm e}^{iJt/4}\bigg[|\nu_1\rangle\langle \nu_1| + \cos\Lambda t\lbrace|\nu_2\rangle\langle \nu_2|+|\nu_3\rangle\langle \nu_3| \rbrace \nonumber \\
&&+ \sin\Lambda t \lbrace|\nu_2\rangle\langle \nu_2|-|\nu_3\rangle\langle \nu_3| \rbrace
+ {\rm e}^{-iJt}|\nu_4\rangle\langle \nu_4| \bigg ].
\end{eqnarray}
where $\Lambda = \sqrt{\Delta_0^2+\omega^2_1}$ and $\Delta_0 = \omega-\omega_0$. The eigen vectors, $|\nu_i\rangle$ and the transformation matrix are given in Appendix A. The above form gets simplified at resonance i.e., for $\omega = \omega_0$.
The unitary operator can then be written, in terms of the two-qubit spin operators as
\begin{eqnarray}
\label{tqu}
\hspace{-10mm}
U_{12}(t) &=&  \hat{V}_{12}{\rm e}^{iJt/4}\bigg[(\frac{1}{4}\mathcal{I} +\vec{S}_1\cdot\vec{S}_2-2S^x_1S^x_2) \nonumber \\
&&+ \cos\omega_1 t(\frac{1}{2}\mathcal{I}+2S^x_1S^x_2)
+ i\sin\omega_1 t (S^x_1 +S^x_2) \nonumber \\
&&+ {\rm e}^{-iJt}(\frac{1}{4}\mathcal{I} - \vec{S}_1\cdot\vec{S}_2) \bigg ].
\end{eqnarray}
Using the above the transition probabilites between the 
states $|\uparrow\uparrow\rangle \leftrightarrow |\downarrow\downarrow\rangle$ and
$|\uparrow\downarrow\rangle \leftrightarrow |\downarrow\uparrow\rangle$ can be easily evaluated. They are given by
\begin{eqnarray}
\label{tqtrans}
P_{\downarrow\downarrow}(t) &=& \frac{1}{2}\left(1-\cos\omega_1 t\right)^2, \nonumber \\
P_{\downarrow\uparrow}(t) &=& \frac{1}{2}\left\vert1-{\rm e}^{iJt}\cos\omega_1 t\right\vert^2. 
\end{eqnarray}

The transitions within the triplet sector i.e., between the states $|\uparrow\uparrow\rangle \rightarrow |\downarrow\downarrow\rangle$, are controlled only by the frequency of the transverse field $\omega_1$. On the otherhand for $|\uparrow\downarrow\rangle \rightarrow |\downarrow\uparrow\rangle$, which invovles both the singlet and the triplet states, the transitions occur both due to the exchange interaction and the time-dependent field. For $\omega_1  \ne J$ there are two frequencies for Rabi oscillations and this would lead to a beat pattern for $P_{\downarrow\uparrow}(t)$. Note that the swapping between the qubits in the case of $|\uparrow\downarrow\rangle \rightarrow |\downarrow\uparrow\rangle$, can be controlled both by $J$ and $\omega$, which can be useful during gate operations.

It is easy to show that 
in the transition from $|\uparrow\uparrow\rangle \rightarrow |\downarrow\downarrow\rangle$,
no entanglement is generated with time. On, the contrary for the transition, $|\uparrow\downarrow\rangle \rightarrow |\downarrow\uparrow\rangle$, entanglement is generated
and it is given by 
\begin{eqnarray}
\mathcal{C}(t) = \frac{1}{2}\left\vert 1-{\rm e}^{iJt}\cos^2\omega_1t - \sin^2\omega_1t\right\vert
\end{eqnarray}
In obtaining the above we have considered the resonant frequency for $\omega$. The entanglement here is mainly due to the exchange interaction between the qubits i.e, if $J=0$, $\mathcal{C}(t) = 0$.  The time-dependent field, can only tune its value, such that the time periods, for which the state becomes maximally entangled and unentangled can be controlled.
We now solve the dynamics in the presence of bath interaction.
\begin{figure}[htb]
   \includegraphics[width=8.0cm]{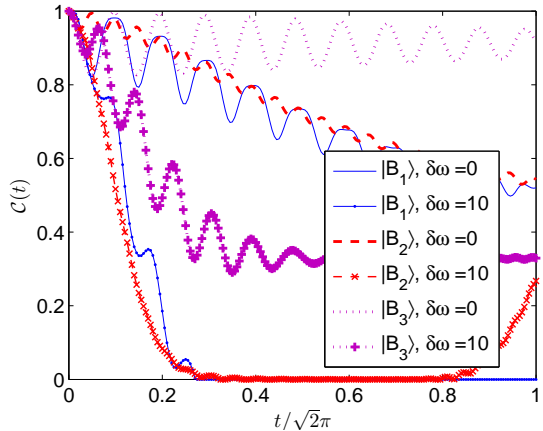}
 \caption{(Color online) Concurrence with time. The time-dependent concurrence, for the Bell-states
  $|B_1\rangle = {1\over \sqrt{2}}[|\uparrow \downarrow \rangle + |\downarrow \uparrow \rangle]$
$|B_2\rangle = \frac{1}{\sqrt{2}}[|\uparrow \uparrow \rangle + |\downarrow \downarrow \rangle]$,$ |B_3\rangle = \frac{1}{\sqrt{2}}[|\uparrow \uparrow \rangle - |\downarrow \downarrow \rangle]$, is plotted for two values of $\Delta \omega = \omega-\omega_0$. The loss of entanglement is slowest at the resonant frequency $\delta \omega=0$, for all the states, and it increases as the r.f frequency shifts away from $\omega_0$. The bath is unpolarized with uniform coupling strengths between the qubits and bath-spins. In plotting the above we have taken
$N=20$, $\omega=100$, $\omega_1 = 10$ and $g/\hbar=1$. The frequencies are in the units of MHz.}
 \end{figure}
\subsubsection{Coupling to N-spin bath}
As shown in Sec-II that the Hamiltonian has a block diagonal structure in the $I^z$ basis of the bath, the unitary operator here can be written using the above obtained form of $U_{12}(t)$ (see Eq.\ref{tqutop} as
\begin{eqnarray}
\label{eqn6}
{U}(t) = \sum_{i=1}^{2^N}{U}_{12}^i|i\rangle\langle i|,
\end{eqnarray}
where $|i\rangle$ are the basis states of the bath and the expression for ${U}_{12}^i$ is same as that given in Eq.\ref{tqu}, except that the Rabi frequency $\Lambda$ gets replaced
by $\Lambda_i = \sqrt{\omega^2_1 + \Delta^2_i}$, where $\Delta_i = \omega-\left[\omega_0 + \langle i|\sum_k g_k I^z_k|i\rangle\right]$.

The reduced density matrix for the two-qubit system can be written as 
\begin{eqnarray}
\label{eqnred}
\rho_S(t) =  \frac{1}{2^N}\sum_{i=1}^{2^N} \lambda_i {U}_i \rho_S(0) {U}^{\dagger}_i 
\end{eqnarray}
where $\lambda_i = (1+P)^{N/2+m_i}(1-P)^{N/2-m_i}$, where $m_i = 2 \langle i |\sum_k I^z_k |i\rangle$.

In the presence of the bath interaction, the width of the distribution for transition probabilities  considered earlier in Eq.\ref{tqtrans} increases, and the amplitude of the Rabi-oscillations between the two-qubit states decreases.
Even these features are same as found in the case of a single qubit, the interest here is the entanglement and its variation with the r.f frequency $\omega$ and the bath polarization $P_B$.
It is important to note that the bath does not induce any entanglement between the initially unentangled states, but if there is any non-zero entanglement at $t=0$ it decays with time.
For the two qubits evolving under their self Hamiltonian ($g_k's=0$), there is no decay of entanglement for any of the maximally entangled states (Bell-states) $|B_1\rangle = {1\over \sqrt{2}}[|\uparrow \downarrow \rangle - |\downarrow \uparrow \rangle]$
$|B_2\rangle = \frac{1}{\sqrt{2}}[|\uparrow \downarrow \rangle + |\downarrow \uparrow \rangle]$,$ |B_3\rangle = \frac{1}{\sqrt{2}}[|\uparrow \uparrow \rangle + |\downarrow \downarrow \rangle]$, eventhough they evolve with time.  
Because of the bath interaction the
two-qubit state becomes mixed with time, and the initial entanglement is no more preserved. 

\begin{figure}[htb]
   \includegraphics[width=8.0cm]{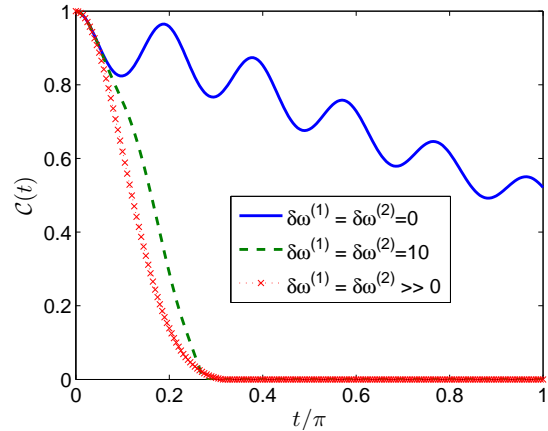}
 \caption{(Color online) Concurrence with time. The time-dependent concurrence for an initial singlet shared between two non-interacting qubits is plotted for three different cases, (i) when the r.f frequencies are tuned to their resonance values i.e., $\delta\omega^{(1,2)} = 0$, (ii)
the r.f frequencies are slightly away from resonance and (iii) much away from their resonant frequencies. In the above $\delta\omega^{(1,2)} = \omega^{(1,2)}_{0}-\omega^{(1,2)}$, where
 $\omega^{(1,2)}_{0}$, are the fields at the local sites of the qubits. The initial states of the baths are unpolarized each consiting of $N=20$ spins and the local fields experienced by the qubits
 are $\omega^{(1)}_{0} = 100$,  $\omega^{(1)}_{0} = 110$ respectively.}
 \end{figure}

In Fig.6 we have plotted the variation of concurrence with time for the entangled states mentioned above. The rate of loss of entanglement is less, when the r.f frequency is tuned to its resonance value ($\omega_0$), and the decay increases as it is tuned away from resonance. 
The dynamics is able to differentiate the evolution of all the Bell-states. 
For non-zero bath polarization the resonant frequency shifts away from $\omega_0$ by $gP_BN/2$, and
this in turn controls the entanglement loss. For example if $g=1$, $P_B=1/2$ and $N=20$, the plots of $\mathcal{C}(t)$ for a given Bell-state in Fig.6 would have been reversed as the condition for resonant frequency has now become $\delta\omega = 10$.  Even though, not shown here, the concurrence and the decoherence have similar initial decay rates. Thereafter the concurrence falls off rapidly to zero. Similar trends are reported \cite{durga2} even with a more general Hamiltonian, for zero external fields.
Note that neither the concurrence nor the decoherence have any dependence on the exchange interaction $J$, as all the states considered above belong to a single spin-sector (triplet).

\subsection{Qubits interacting with separate spin-baths}
In this section we consider the other extreme, where the bath-spins with which the qubits are interacting, are different. This corresponds to a very weak overlap between the electronic wave-functions in the case of a double dot system.
Here, the qubit-bath interaction term does not commute with the exchange term. 
This causes transitions between the singlet and triplet subspaces, due to which the singlet state also decohers. 
The value of exchange interaction will be small in comparison to the common spin-bath case, as $J$ is a direct measure of the overlap function. We first consider the evolution entangled states shared between two remote (non-interacting) qubits and then switch on $J$ to see its effect on the evolution.

\subsubsection{J=0}
When there is no exchange interaction, the two qubits evolve independent of each other, and an initial direct product state will evolve into a direct product
state for later times, i.e.
if $\rho_{S}(0)=\rho_1(0)\otimes\rho_2(0)$, then we have $\rho_{S}(t)=\rho_1(t)\otimes\rho_2(t)$.
Hence the dynamics is of interest only when the initial state is entangled.
This can be a situation where an entangled state is shared between two remote qubits and each of the qubit state is decohering because of their interaction with their individual baths.

For a Bell state we have $\vec P_1 =\vec P_2 =0$, (see Eq.\ref{tqs}) and only the 
diagonal components of $\Pi$ to be non vanishing. The two-qubit reduces to form
\begin{eqnarray}
\label{bellst}
\rho_{12} = \frac{1}{4}[\mathcal{I}+4\sum_m \Pi^{mm}{S}^m_1 {S}^m_2].
\end{eqnarray}
Since the time-evolution operator for the present case is just the direct product of that of individual qubits, the time-dependent polarizations
\begin{eqnarray}
\Pi^{mn}(t)\equiv \frac{1}{2^{N-2}}\sum_{i=1}^{2^N} \lambda_i{\rm Tr}[\rho_12(0)S^m_1(t)S^n_2(t)],
\end{eqnarray}
where
\begin{eqnarray}
\label{etam}
S^m_1(t)S^n_2(t) = \sum_i\eta^{(1)}_{mi}(t){S}^i_1\sum_j\eta^{(2)}_{nj}(t){S}^j_2
.\end{eqnarray}
The time-dependent matrices ${\eta}^{(1,2)}$ are given in Appendix B. Note that all components of the tensor polarization become non-zero at a later time. In the absence of external fields,
only the diagonal components change in time, and hence the state at any later time can still be written in the above given form (see Eq.\ref{bellst}). Due to this, the expressions for concurrence and decoherence at a later time, can be easily obtained in terms of the tensor polarizations \cite{durga2}. But,
in the present case, since all $\Pi^{mn}(t)$ are non-zero, an expression for concurrence $C(t)$ in terms of the tensor polarizations is not straight forward.

In Fig.7 we have plotted the concurrence when the two-qubit state is initially a singlet. We have considered different cases, where the r.f frequency at each qubit site has different values. 
One can see that for the resonant frequencies at local sites the loss of entanglement is minimum. As the r.f frequency moves away from resonance the entanglement falls off rapidly. Eventhough, the initial loss is same for all the cases, the time at which the initial singlet completely looses it entanglement depends on the local oscillating frequency. For frequencies much away from resonance, there is a limiting curve, as shown in Fig.7.
All other Bell-states show the same behavior indicating that there is no special choice for the maximally entangled states when the two-qubits are non-interacting.

For intermediate regimes where the exchange interaction is neither strong nor weak, we have few number of bath-spins which are common to both and the baths can be still be assumed to be separate. As the triplet-singlet separation is decided by the value of $J$, the effect of non-zero $J$ for separate baths is to protect the singlet state from the effects of decoherence. Any non-zero $J$ would differentiate the behavior of $\mathcal{C}(t)$ for all the Bell-states.

\section{Conclusion}
In this work we have studied an exactly solvable model for the dynamics of one and two-qubit states interacting with a spin-bath under the influence of a time-dependent magnetic field. 
The magnetic field is static along $\hat{z}$ direction and is oscillatory ($r.f$) in the transverse plane ($\hat{x}-\hat{y}$).
The expressions for the reduced density matrices of the qubit states at a later time have been obtained. Various distributions for the qubit-bath couplings are considered. In the case of single-qubit the transition probability and Rabi-oscillations between any two orthogonal states, depends on the strength and distribution of qubit-bath couplings and initial bath polarization. The strength of qubit-bath interaction, determines the peak value and width of the distribution for transition probability, whereas the resonant frequency for the transition is decided by the bath-polarization. There is a slower power-law decay ($t^{3/2}$) for the Rabi-oscillations for long times, in contrast to a fast Gaussian decay for small times. In the case of two interacting qubits, the presence of a $r.f$ field, gives an extra tunable parameter in addition to the exchange interaction, for the swapping of quantum states between the qubits. Under collective decoherence (common spin-bath) the loss of entanglement is different for all the Bell-states, whereas, for the two qubits which are non-interacting and decohering due to their individual environments, all Bell-states decoher at the same rate. By tuning the $r.f$ frequency the rate of loss of entanglement can be controlled in both the cases.

\appendix
\section{Eigen states of the two-qubit Hamiltonian}
The eigen vectors for the two-qubit Hamiltonian in the absence of bath interaction (see Eq.\ref{tqs}) are given by
\begin{eqnarray}
|\nu_1\rangle &=& |1\rangle_T -\frac{\sqrt{2}\Delta}{\omega_1}|0\rangle_T - |-1\rangle_T \nonumber \\
|\nu_2\rangle &=& \left(1-2\Lambda\frac{\Lambda-\Delta}{\omega^2_1}\right)|1\rangle_T -\frac{\sqrt{2}(\Lambda-\Delta)}{\omega_1}|0\rangle_T - |-1\rangle_T \nonumber \\
|\nu_3\rangle &=& \left(1-2\Lambda\frac{\Lambda+\Delta}{\omega^2_1}\right)|1\rangle_T -\frac{\sqrt{2}(\Lambda+\Delta)}{\omega_1}|0\rangle_T - |-1\rangle_T \nonumber \\
|\nu_4\rangle &=& |0\rangle_S
\end{eqnarray}
where $\Lambda = \sqrt{\Delta^2+\omega^2_1}$ and $\Delta=\omega-\omega_0$. The above eigenvectors are unnormalized and written in the total-spin basis, where the basis vectors
$|1\rangle_T = |\uparrow\uparrow\rangle$, $|-1\rangle_T = |\downarrow\downarrow\rangle$,
$|0\rangle_T = \frac{1}{\sqrt{2}}[|\uparrow\downarrow\rangle + |\downarrow\uparrow\rangle]$,
and $|0\rangle_S = \frac{1}{\sqrt{2}}[|\uparrow\downarrow\rangle - |\downarrow\uparrow\rangle]$.
The above eigen-vectors are used in constructing the unitary operator Eq.\ref{tqutop}. 

The transformation matrix $\hat{V}_{12}$ (see Eq.\ref{tqutop}) connecting the rotating frame and lab frame is given by
\begin{eqnarray}
\hat{V}_{12} = \left[\begin{array}{cccc}{\rm e}^{i\omega t} & 0 & 0 & 0 \\
                              0 & 1 & 0 & 0 \\
                              0 & 0 & {\rm e}^{-i\omega t} & 0 \\
			      0 & 0 & 0 & 1\end{array}\right ].
\end{eqnarray}
The matrix is written in the total spin basis of the two-qubits given above.

\section{Time-dependent matrices (${\eta}^{(1,2)}$)}
The time-dependent coefficents $\eta^{(1)}_{mi}(t)$ appearing in Eq.\ref{etam} are given by
\begin{widetext}
\begin{eqnarray}
\hat{\eta}^{(1)} = \hat{R}\frac{1}{\Omega^2_i}
 \left[\begin{array}{ccc}
 \left(\omega^2_1 + \Delta^2_i\cos \Omega_i t\right) & -\Delta_i\Omega_i\sin \Omega_i t &-2\Delta_i\omega_1\sin^2\frac{t}{2}\Omega_i  \\
                              \Delta_i\Omega_i \sin \Omega_i t & \Omega^2_i \cos\Omega_i t & -\omega_1\Omega_i\sin\Omega_i t \\
                               2\Delta_i\omega_1\sin^2\frac{t}{2}\Omega_i &
                                \omega_1\Omega_i\sin \Omega_i t & 
                              \left(\Delta^2_i + \omega^2\cos\Omega_i t\right)\end{array}\right ]\hat{R}^\dagger
\end{eqnarray}
\end{widetext}
where $\Omega_i = \sqrt{\Delta^2_i + \omega^2_1}$ and $\Delta_i = \omega^{(1)}-\omega^{(1)}_0$. In a similar way the matrix $\hat{\eta}^{(2)}$ can be written with $\Delta_i = \omega^{(2)}-\omega^{(2)}_0$. Here $\omega^{(1,2)}_0$ represents the local fields at the qubit sites and $\omega^{(1,2)}$, the respective local r.f frequencies.
The transformation matrix $\hat{R} = \cos({\omega^{(1)}\frac{t}{2}})\hat{\mathcal{I}} + 2i\sin({\omega^{(1)}\frac{t}{2}}) S^z_1$.
\begin{acknowledgments}
The author would like to thank Sunil Kr. Mishra for his assistance while preparing the manuscript.
\end{acknowledgments}

\end{document}